\documentclass[conference,a4paper]{APSIPA2021}
\IEEEoverridecommandlockouts
\usepackage{amsmath}
\usepackage{multirow}
\usepackage{threeparttable}
\usepackage{diagbox}
\usepackage{romannum}
\usepackage{amssymb}
\usepackage{subcaption}

\usepackage{ifpdf}
\ifpdf
  \usepackage[pdftex]{graphicx}
\else
  \usepackage[dvipdfmx]{graphicx}
\fi

\usepackage{cite}
\usepackage{url}
\usepackage{cleveref}
\crefname{figure}{Fig.}{Figs.}
\Crefname{figure}{Fig.}{Figs.}

\usepackage{geometry}
\usepackage{fancyhdr}

\fancypagestyle{firststyle}{
  \fancyhf{}
  \fancyhead[C]{2026 Asia Pacific Signal and Information Processing Association Annual Summit and Conference (APSIPA ASC)}
}

\begin{document}

\title{Enhancing Privacy in Federated Learning via \\ Dual Obfuscation of Gradients and Training Images}

\author{
\authorblockN{
Yuki Itabashi\authorrefmark{1},
Hiroto Sawada\authorrefmark{1}, 
Mare Hirose\authorrefmark{1}, 
Shoko Imaizumi\authorrefmark{1}, and 
Hitoshi Kiya\authorrefmark{2}
}
    \authorblockA{
        \authorrefmark{1}Chiba University, Chiba, Japan\\
        \authorrefmark{2}Tokyo Metropolitan University, Tokyo, Japan\\
        Email: 26wm3205@student.gs.chiba-u.jp, \{marehirose, imaizumi\}@chiba-u.jp, kiya@tmu.ac.jp 
}
\thanks{This work was partially supported by JSPS KAKENHI Grant Number JP25K07750.}
}
\maketitle
\thispagestyle{firststyle}
\pagestyle{fancy}

\begin{abstract}

Federated learning enables collaborative model training while keeping data locally at each client; however, recent studies have shown that training data can be reconstructed from shared model updates. To address this issue, this paper proposes a dual obfuscation method that enhances robustness against image restoration attacks by jointly obfuscating updated information and training images. The proposed method combines a robustness enhancement technique based on random binary weights, which randomly sets a portion of gradient elements to zero, with an image encryption technique. These techniques provide complementary protection by reducing the amount of original gradient information available to an attacker and the visual interpretability of reconstructed images, respectively. 
Furthermore, the image encryption technique allows independent keys to be used for each client and each image, avoiding explicit key sharing. 
Experimental results on an image classification task using a Vision Transformer (ViT) show that the proposed method reduces the visual information recovered by Attention Privacy Leakage (APRIL) under the evaluated settings without causing additional degradation in classification performance beyond that caused by image encryption. 
Although the proposed combination does not provide an absolute security guarantee, the results demonstrate the potential benefit of combining gradient modification and image encryption for privacy-enhanced federated learning.

\end{abstract}

\pagestyle{empty}

\raggedbottom
\section{Introduction}
In recent years, developments in AI technology have led to an increasing number of opportunities to use deep learning. While training high-performance models requires a large amount of data, collecting such data demands a lot of time and effort, and ensuring security in data management has become an important challenge. Federated learning has attracted much attention as an effective approach to addressing these issues \cite{Fed_file}.
\par
\begin{figure}[tb]
\begin{center}
\includegraphics[width=1.0\linewidth]{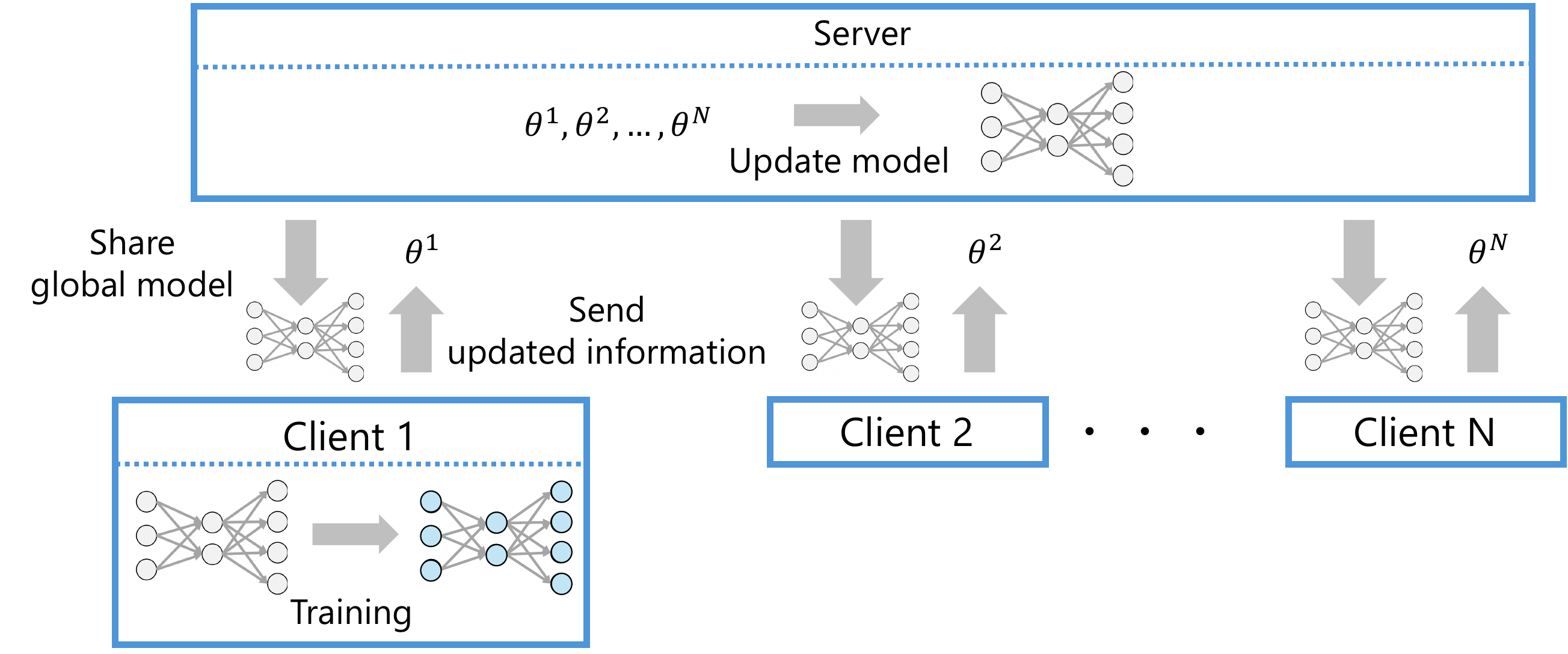}
\end{center}
\caption{Overview of federated learning.}
\label{fig:Abstract_of_FL}
\end{figure}

As shown in Fig. \ref{fig:Abstract_of_FL}, federated learning is a framework in which each client retains its local training data while only updated model information is shared for training. With this characteristic, federated learning is expected to be an effective method for privacy preservation. However, recent studies have reported attacks that restore training data from the shared updated information \cite{Attack_Membership, Attack_DeepLearningBase, Attack_HighRestore_formBatch, APRIL}, making the enhancement of security in federated learning a critical challenge. In this paper, we address this issue using an image classification task as an example.
\par
To address this issue, several security enhancement methods that obfuscate the updated information shared with the server have been proposed. Representative approaches include differential privacy, which adds noise to the updated information \cite{DP_Fed, DP_communication_cost, DP_Layer}, and methods that replace part of the updated information with arbitrary values \cite{Attack_DeepLearningBase, APRIL, RBW_sawada_file}. These methods enhance robustness against image restoration attacks that exploit the shared updates; however, issues remain regarding the determination of the appropriate strength of enhancement and robustness against more advanced attacks.
\par
As another approach for privacy preservation, image encryption techniques that obfuscate the visual information of training images have been proposed \cite{ImageEnc_Fed, ImageEnc_Aso, ImageEnc_ShareGAN}. When these techniques are applied to federated learning, the visual information remains obfuscated even in images restored from updated information. However, many image encryption techniques require sharing encryption keys among clients, which poses risks such as the leakage of encryption keys or unauthorized use by other clients. Therefore, challenges remain from the viewpoint of privacy preservation.
\par
To tackle the above issues, we propose a security enhancement method that provides dual obfuscation for client privacy by simultaneously applying a conventional robustness enhancement technique and an image encryption technique. The two techniques address different aspects of information leakage and are used in a complementary manner. Their combination is intended to reduce the risk of visually interpretable training images being reconstructed rather than to provide an absolute security guarantee. Furthermore, the image encryption used in this study allows the use of different keys for each client or each image, which improves security and mitigates the impact of attacks. In our experiments, we use the Vision Transformer (ViT) \cite{vit_file} as an example and apply federated learning using randomly selected model parameters (FLRSP) \cite{RBW_sawada_file}, a robustness enhancement technique that randomly replaces part of updated information with zero values, together with an image encryption technique that has high compatibility with ViT \cite{encryption_hirose}. We evaluate the effectiveness of the proposed method in terms of robustness against image restoration attacks and model performance.

\section{Preparation}
In this paper, we propose a method to enhance robustness against image restoration attacks in federated learning, using ViT as an example. In this section, we first introduce the basic architecture of ViT, followed by a review of previous studies on attack methods and robustness enhancement techniques in federated learning.

\subsection{Vision Transformer}\label{ViT}
\begin{figure}[tb]
\begin{center}
\includegraphics[width=1.0\linewidth]{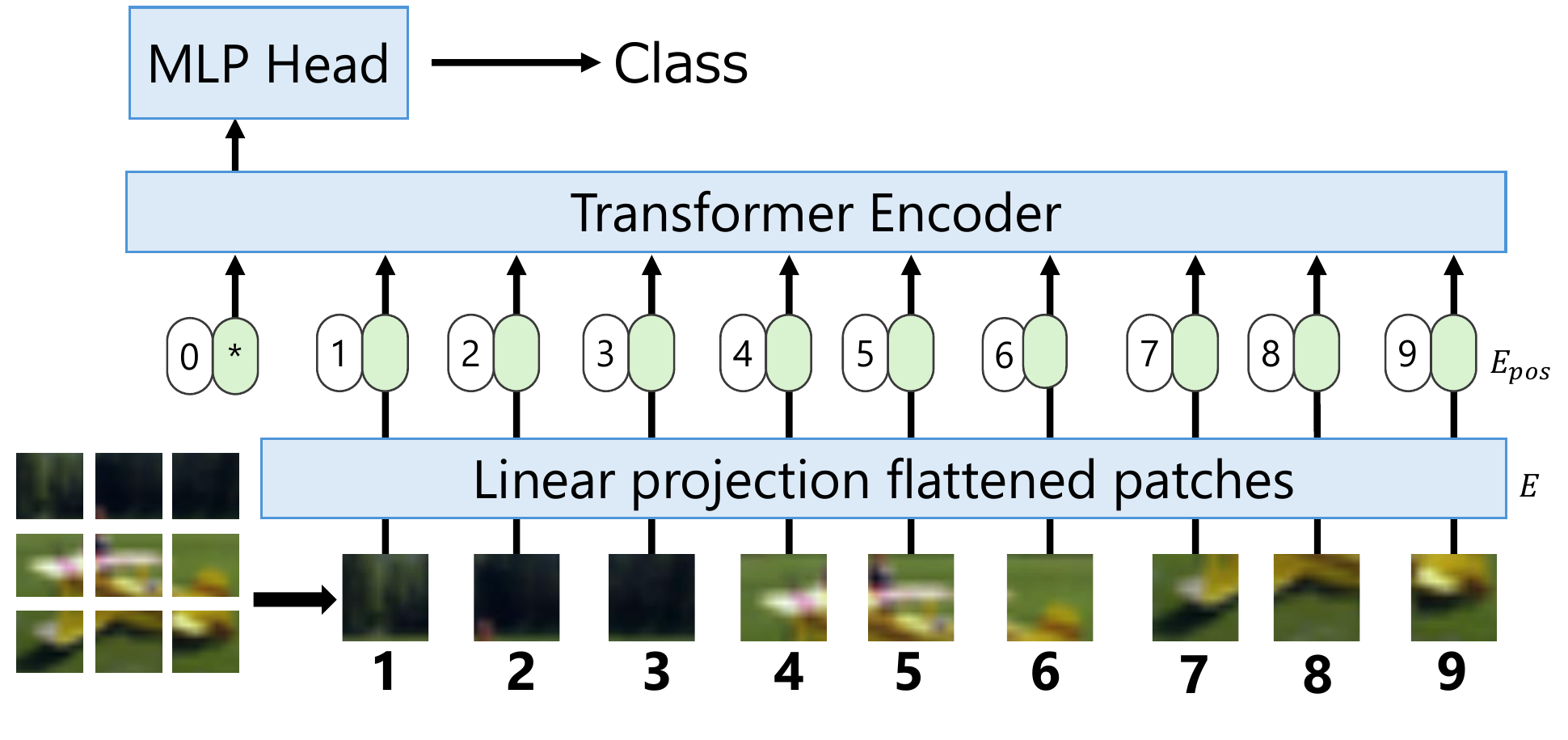}
\end{center}
\caption{ViT structure.}
\label{fig:Abstract_of_ViT}
\end{figure}

ViT \cite{vit_file} is an image classification model based on attention mechanisms and is known for achieving a high classification  performance.  Fig. \ref{fig:Abstract_of_ViT} illustrates the structure of ViT. First, an input image is divided into small regions called patches, which are then transformed into the input dimension of the Transformer Encoder by a linear layer denoted as $E$. Next, a class token representing the class information is appended to the beginning of the patch sequence, and positional embeddings are added to each patch to encode positional information. The resulting patch sequence is then fed into the Transformer Encoder layers, and finally, only the class token is input to the MLP Head to predict the class label of the input image.

\subsection{Related Work}
Federated learning is a distributed learning framework in which a global model is trained by aggregating updated information obtained from multiple clients through local training. Since raw training data are not shared among clients or with the server, federated learning is expected to provide privacy preservation. However, recent studies have demonstrated that sensitive information can still be inferred from shared updated information, which can lead to potential privacy leakage.
\par
Several image restoration attacks have been proposed to reconstruct training data from shared updated information \cite{Attack_Membership, Attack_DeepLearningBase, Attack_HighRestore_formBatch, APRIL}. Typical approaches include membership inference attacks \cite{Attack_Membership}, which estimate whether specific data were used during training, and optimization-based reconstruction attacks \cite{Attack_DeepLearningBase}, which iteratively recover input images by minimizing the difference between gradients. These attacks exploit the structural properties of model updates, and they indicate that the direct sharing of updated information poses a significant privacy risk.
\par
To mitigate such attacks, various robustness enhancement techniques have been proposed. Differential privacy is a representative approach that adds stochastic noise to updated information before sharing it with the server \cite{ DP_Fed, DP_communication_cost, DP_Layer}. While this technique provides formal privacy guarantees, it introduces a trade-off between privacy protection and model performance; stronger noise generally leads to a degradation in model accuracy.

\par
As an alternative approach, techniques that partially replace updated information with arbitrary values have been investigated \cite{Attack_DeepLearningBase,APRIL,RBW_sawada_file}. In particular, randomly replacing a portion of updated information with zero values has been shown to effectively reduce the risk of image restoration attacks such as Attention Privacy Leakage (APRIL) \cite{RBW_sawada_file}. This technique maintains model performance by avoiding constraints on specific layers and selectively modifying the updated information. However, since the remaining updated information still preserves its original structure, attackers can exploit this information to reconstruct training images.

\section{Proposed Method}
In this paper, we propose a method that enhances robustness against image restoration attacks in federated learning by combining a robustness enhancement technique with an image encryption technique. The proposed method provides dual obfuscation by modifying both the updated information and the visual information of training images. The two techniques play complementary roles and should not be regarded as complete backups for each other. Their combination does not guarantee complete protection, but it may reduce the reconstruction risk when either technique alone provides insufficient protection.

\subsection{Overview of Proposed Method}\label{section3_abst}
\begin{figure}[tb]
\begin{center}
\includegraphics[width=0.9\linewidth]{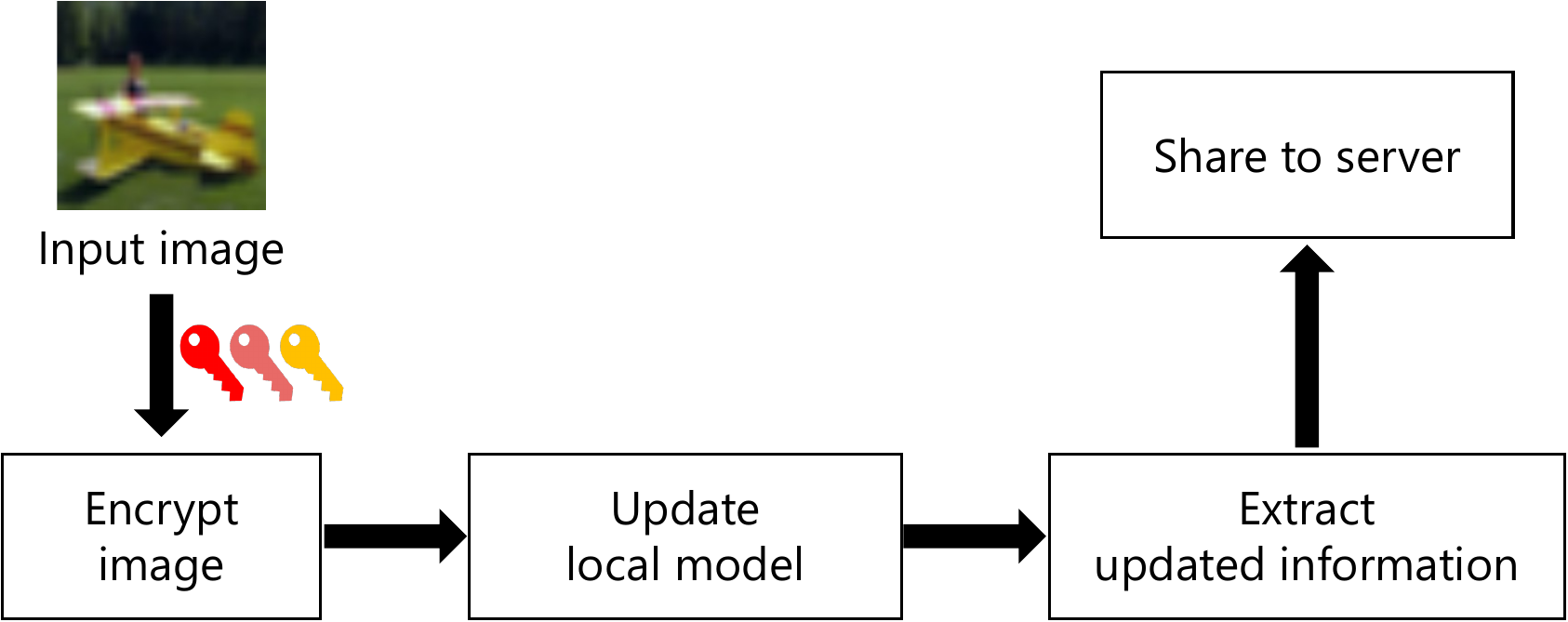}
\end{center}
\caption{Processing procedure in each client.}
\label{fig:clinet_doing}
\end{figure}

Fig. \ref{fig:clinet_doing} shows an overview of the proposed method. Each client first encrypts its local training images using an image encryption technique before performing local training. The encrypted images are then used as input to the model, and the model is trained using these encrypted images.
\par
After local training, each client generates updated information on the basis of the encrypted images. Subsequently, a robustness enhancement technique is applied to the updated information by partially replacing it with zero values \cite{RBW_sawada_file}. The processed updated information is then transmitted to the server for aggregation.
\par
Because the updated information and the visual content of the training images are obfuscated through different operations, their combination can reduce the amount of interpretable information recovered by an attacker. The contribution of each operation depends on the values of $R_W$ and $R_E$. Therefore, the two operations should be regarded as complementary mechanisms rather than complete backups for each other, and their combination does not guarantee protection against all reconstruction attacks.

\subsection{Image Encryption}\label{image_Enc}

\begin{figure}[tb]
\begin{center}
\includegraphics[width=0.9\linewidth]{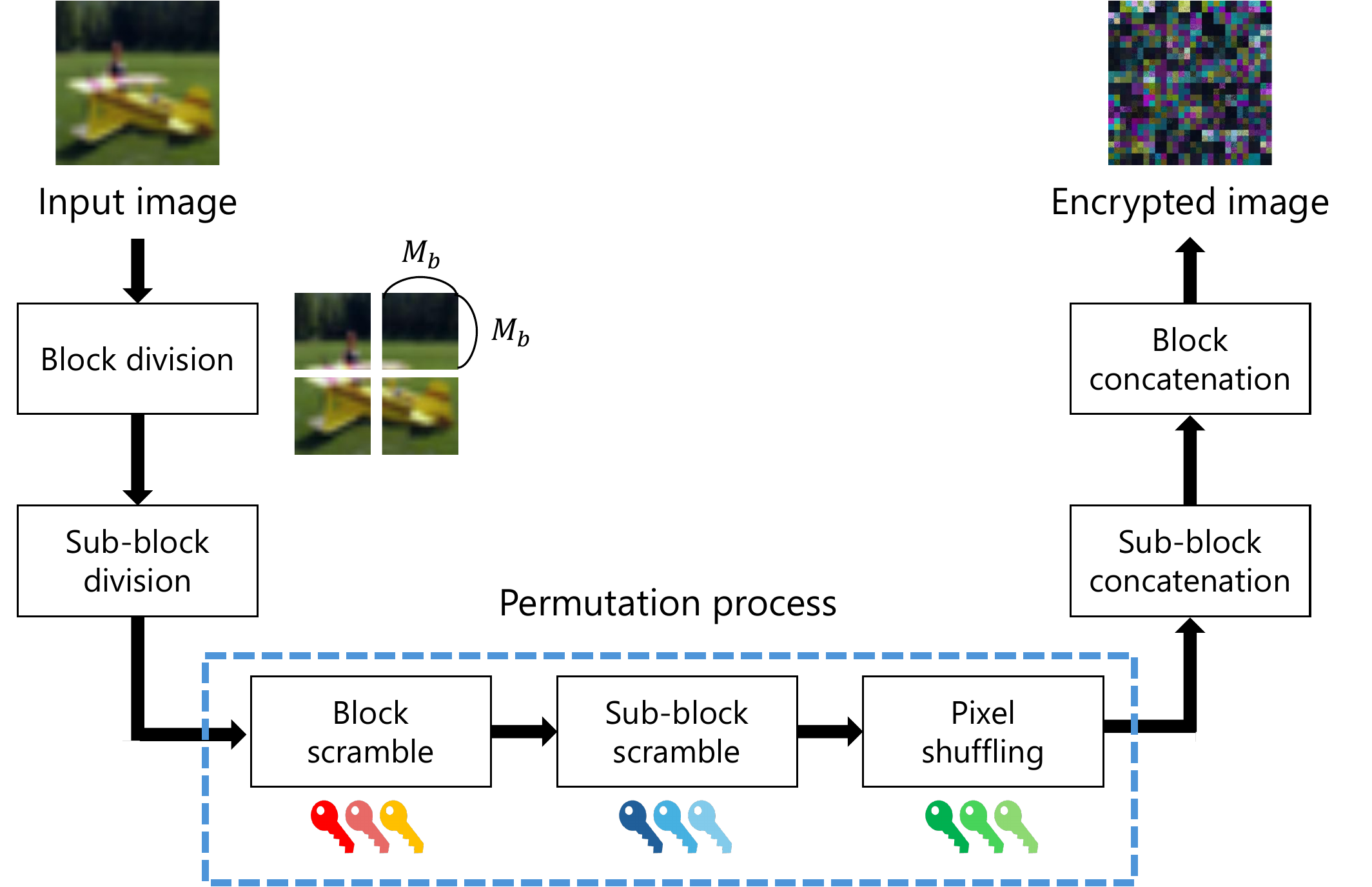}
\end{center}
\caption{Procedure for generating encrypted images.}
\label{fig:Abst_Image_Enc_Block}
\end{figure}

\begin{figure}[tb]
\begin{center}
\includegraphics[width=0.7\linewidth]{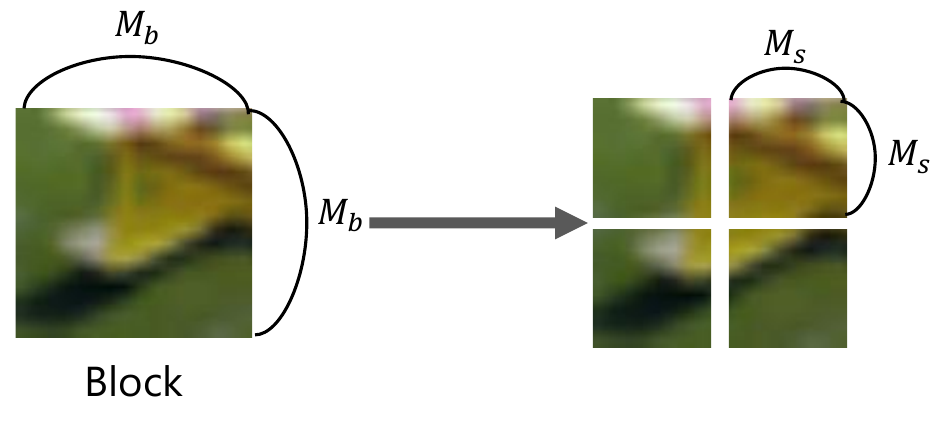}
\end{center}
\caption{Division of block into sub-blocks.}
\label{fig:Block_to_SubBlock}
\end{figure}

\begin{figure}[tb]
\begin{center}
\includegraphics[width=1.0\linewidth]{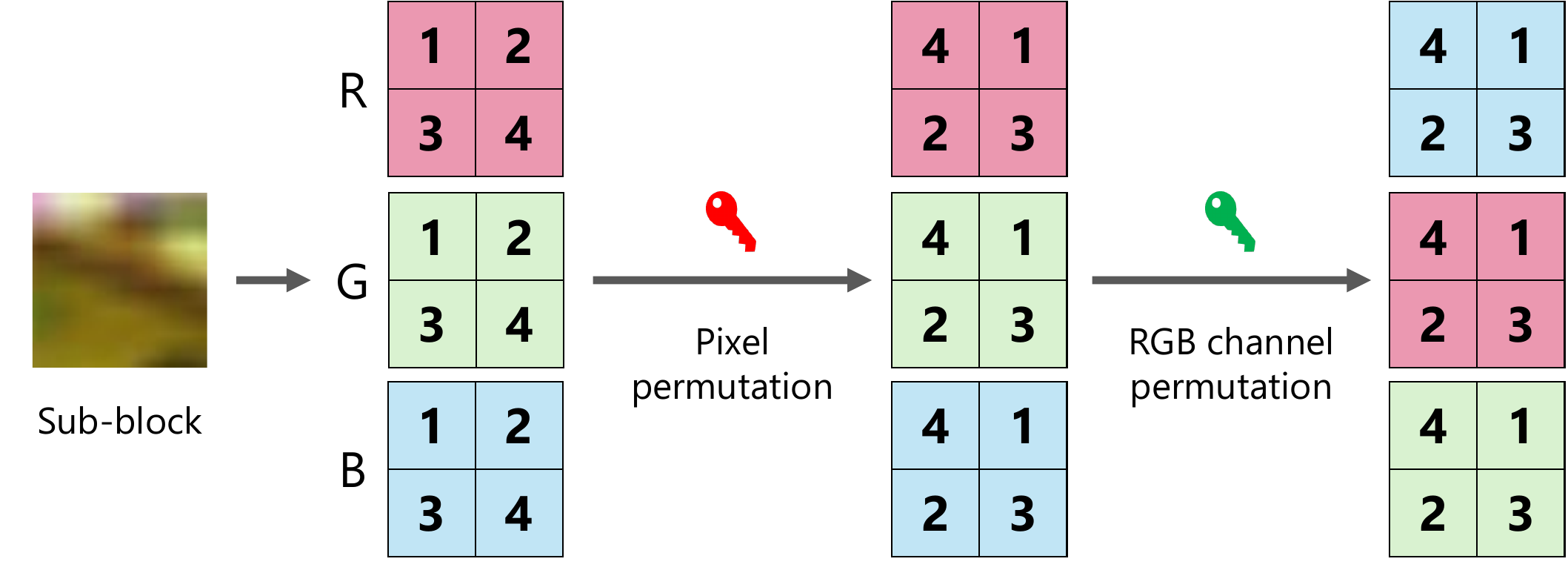}
\end{center}
\caption{Encryption process within each sub-block.}
\label{fig:Pixel_Shuffle}
\end{figure}

As described in Section \ref{section3_abst}, the proposed method applies an image encryption technique to input images in order to obfuscate their visual information. The overall procedure of the encryption process is illustrated in Fig. \ref{fig:Abst_Image_Enc_Block}.
\par
In this study, we adopt an image encryption technique reported to have high compatibility with ViT \cite{encryption_hirose}. The detailed procedure is described as follows (see \Cref{fig:Abst_Image_Enc_Block,fig:Block_to_SubBlock,fig:Pixel_Shuffle}).
\begin{quote}\renewcommand{\labelenumi}{\bf Step \theenumi:}
\begin{enumerate}{
\item Each client divides an input image $x$ into non-overlapping blocks of size $M_b \times M_b$. A random integer sequence generated using a secret key is used to permute the positions of the blocks (block scrambling). A different key can be used for each client and each image.
\item Each block is further divided into sub-blocks of size $M_s \times M_s$, as shown in Fig. \ref{fig:Block_to_SubBlock}. The positions of the sub-blocks within each block are randomly permuted (sub-block scrambling). A different key can be used for each block.
\item Pixel shuffling is performed within each sub-block. As illustrated in Fig. \ref{fig:Pixel_Shuffle}, this process consists of two steps: permutation of pixel positions and permutation of RGB channels within each sub-block.
}
\end{enumerate}\end{quote}
Note that it is not necessary to apply the permutation process to all blocks in the image. In block scrambling, let $N_{\text{total}}$ be the total number of blocks and $N_{\text{select}}$ be the number of blocks selected for permutation. Then, the ratio of permuted blocks $R_E$ is given by
\begin{equation}
R_E = \frac{N_{\text{select}}}{N_{\text{total}}}.
\label{eq:RE_frac}
\end{equation}
When $N_{\text{select}} = 0$, $R_E$ is 0.0, and no block permutation is performed; thus, the original image is preserved. In contrast, when $N_{\text{select}} = N_{\text{total}}$, $R_E$ is 1.0, and all blocks are subject to encryption. Furthermore, $R_E$ can be defined in the same manner for sub-block scrambling and pixel shuffling within each sub-block. When $R_E < 1.0$, some blocks are not processed by the corresponding permutation operation. Consequently, local visual information may remain in these blocks and could be recovered through an image reconstruction attack. Therefore, $R_E=1.0$ is preferable when visual privacy is prioritized, although increasing $R_E$ may reduce classification accuracy, as shown in Section \ref{Result_Acc}.

\subsection{Local Model Update and Extraction of Updated Information}\label{model_local}
In the proposed method, each client performs local training using encrypted images generated by the image encryption technique described in Section \ref{image_Enc}. In contrast to conventional methods that directly utilize plain training images, the proposed method updates the local model using encrypted images. As a result, the visual information of the original training data is not exposed.
\par
After local training, each client extracts updated information from the trained model. Let $\theta_{A,i}^{n,m}$ denote the updated information corresponding to the $i$-th parameter in layer $A$ at iteration $m$, obtained from client $n$. This updated information is computed on the basis of the encrypted images used during training.
\par
The updated information is derived from encrypted images. Therefore, even if an attacker attempts to reconstruct training data, the results correspond to encrypted representations rather than the original images. This property contributes to preventing the leakage of visually interpretable information.

\subsection{Sharing Parameters and Updating Global Model}\label{model_Enc}
In federated learning, the server aggregates the updated information shared by the clients, and the global model is updated on the basis of the following equation:
\begin{equation}
w_{A,i}^{m+1} = w_{A,i}^{m} - \eta \frac{\sum_{n=1}^{N} \theta_{A,i}^{n,m}}{N},
\label{eq:grads_renew}
\end{equation}
where $w_{A,i}^{m}$ denotes the $i$-th parameter in layer $A$ at iteration $m$, and $\theta_{A,i}^{n,m}$ denotes the updated information obtained from client $n$. Furthermore, $\eta$ is the learning rate, and $N$ is the number of clients.

\par
In the proposed method, the updated information shared with the server is computed from encrypted images, as described in Section \ref{model_local}. In addition, the conventional Federated Stochastic Gradient Descent (FedSGD) \cite{Fed_file} is used for model training, and FLRSP, a gradient obfuscation technique based on random binary weights \cite{RBW_sawada_file}, is applied by modifying the updated information before sharing.
\par
\begin{figure}[tb]
\begin{center}
\includegraphics[width=0.9\linewidth]{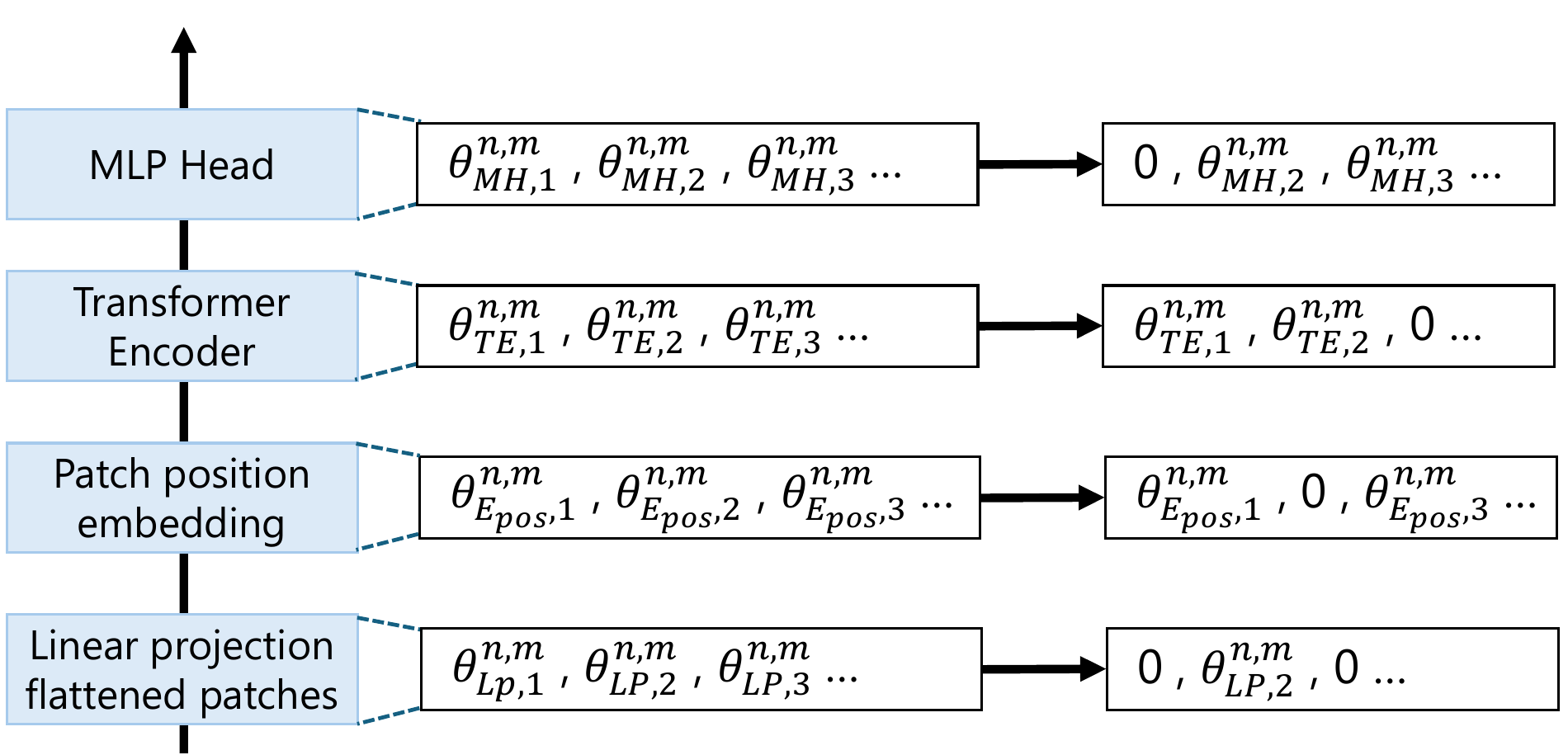}
\end{center}
\caption{Replacement of gradients with random binary weights.}
\label{fig:Abstract_of_RBW}
\end{figure}
As shown in Fig. \ref{fig:Abstract_of_RBW}, FLRSP replaces a part of the gradients with zero values. Let $\theta_{A,i}^{n,m}$ denote the original gradient and $B_{A,i}^{n,m} \in \{0, 1\}$ denote the corresponding random binary weight. Then, the modified gradient $\theta_{A,i}^{n,m'}$ is given by
\begin{equation}
\theta_{A,i}^{n,m'}
= B_{A,i}^{n,m} \times \theta_{A,i}^{n,m} .
\label{eq:theta_changed}
\end{equation}
The weights $B_{A,i}^{n,m}$ are generated independently for each client. Increasing the probability $R_W$ ($0.0 \leq R_W \leq 1.0$) of assigning zero values increases the number of gradients replaced with zero values. While this improves robustness against image restoration attacks, it may degrade model performance because fewer gradients contribute to the update.

\par
Considering that some gradients are replaced with zero values, the global model is updated by aggregating only the non-zero gradients as follows:
\begin{equation}
w_{A,i}^{m+1}
=
\begin{cases}
  w_{A,i}^{m} - \eta \frac{\sum_{n=1}^{N} \theta_{A,i}^{n,m'}}{\sum_{n=1}^{N} B_{A,i}^{n,m}} & \text{if } 0 < \sum_{n=1}^{N} B_{A,i}^{n,m} \leq N , \\
  w_{A,i}^{m} & \text{if }\sum_{n=1}^{N} B_{A,i}^{n,m} = 0 .
\end{cases}
\label{eq:grads_RBW}
\end{equation}
If all gradients are replaced with zero values, the corresponding parameter is not updated.


\section{Features of Proposed Method}
The key features of the proposed method are summarized as follows:
\begin{itemize}
\item FLRSP and image encryption obfuscate different types of information. The former reduces the amount of original gradient information available to an attacker, whereas the latter reduces the visual interpretability of reconstructed images. Their combination may therefore reduce the reconstruction risk when either technique alone provides insufficient protection. However, neither technique can fully compensate for the failure of the other, and their combination does not guarantee protection against all reconstruction attacks.

\item Because independent keys can be used for each image and each client, even if an attacker obtains the encryption key for a particular image, the images that can be restored are limited to that image.

\item The proposed method can be combined with various existing robustness enhancement techniques.

\item The proposed method does not require explicit key management or key sharing among clients.
\end{itemize}

\section{Experimental Results}
In this section, we evaluate the effectiveness of the proposed method in terms of its robustness against image restoration attacks and the classification performance of the model in an image classification task using ViT.

\subsection{Setup}

In this experiment, a server and multiple clients were placed on the same machine, and federated learning based on FedSGD was conducted. Compared with  Federated Averaging (FedAvg), which performs multiple local update steps before aggregation, FedSGD provides a more direct setting for sharing and modifying gradients. FedSGD was adopted because the APRIL attack considered in this study is designed to exploit gradients shared by clients, and the random binary-weight technique directly modifies these gradients before aggregation. Therefore, FedSGD provides a direct setting for evaluating the effect of gradient modification on image reconstruction. The number of clients was set to $N=5$.
As the learning model, we used a ViT model, vit\_small\_patch16\_224 \cite{vit_model16}, which was pre-trained on ImageNet-21k. The patch size of this model is 16. The CIFAR-10 dataset, consisting of 50,000 training images and 10,000 test images, was used as the training dataset. These images were resized to $224 \times 224 \times 3$, which is the input size of ViT, using bicubic interpolation.
\par

In the proposed method combining FLRSP with image encryption, a portion of the gradients was randomly replaced with zero values. The probability of generating zeros, $R_W$, can be independently determined for each client; however, to simplify the discussion, a common $R_W$ was used for all clients. The positions of the gradients to be replaced were varied for each client and each epoch. In this paper, $R_W$ was set to 0.2, 0.5, and 0.8. In the image encryption process, the block size $M_b$ was set to 16 to fit the patch size of ViT, and the sub-block size $M_s$ was set to 8. Independent keys were used for each client and each image, as well as for each block and sub-block, and these keys were updated every epoch. Furthermore, the ratio of blocks subject to permutation, $R_E$, can be independently determined for each permutation process. However, for simplicity of discussion, a common $R_E$ was used and set to 0.5 or 1.0.



\subsection{Robustness Against Image Restoration Attacks}\label{Attack}
In this section, we conducted an image restoration attack using APRIL \cite{APRIL} on 16 training images selected from the CIFAR-10 dataset. The robustness against the attack was evaluated under three conditions: (\romannum{1}) applying only FLRSP \cite{RBW_sawada_file}, (\romannum{2}) applying only image encryption \cite{encryption_hirose}, and (\romannum{3}) applying the proposed method that combines both.

\begin{figure}[tb]
\centering
\captionsetup{justification=centering}
\newcommand{\unifiedsize}{1.67cm}
\begin{subfigure}{0.32\linewidth}
\centering
\includegraphics[width=\unifiedsize,height=\unifiedsize]{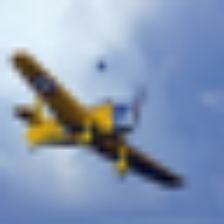}
\caption{Training image\\ (without encryption)}
\label{fig:img_original_line}
\end{subfigure}
\begin{subfigure}{0.32\linewidth}
\centering
\includegraphics[width=\unifiedsize,height=\unifiedsize]{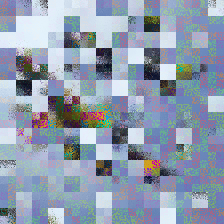}
\caption{Encrypted image\\ ($R_E=0.5$)}
\label{fig:img_05_original_line}
\end{subfigure}
\begin{subfigure}{0.32\linewidth}
\centering
\includegraphics[width=\unifiedsize,height=\unifiedsize]{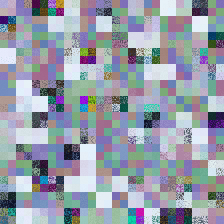}
\caption{Encrypted image\\ ($R_E=1.0$)}
\label{fig:img_10_original_line}
\end{subfigure}

\vspace{4mm}

\begin{subfigure}{\linewidth}
\centering
\renewcommand{\arraystretch}{1.1}

\begin{tabular}{c|c|c|c}
\diagbox{$R_W$}{$R_E$}
 & $0.0$ & $0.5$ & $1.0$ \\
\hline
$0.0$ &
\includegraphics[width=\unifiedsize]{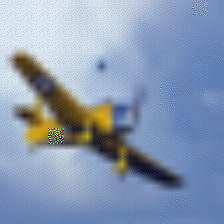} &
\includegraphics[width=\unifiedsize]{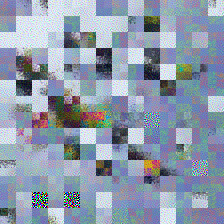} &
\includegraphics[width=\unifiedsize]{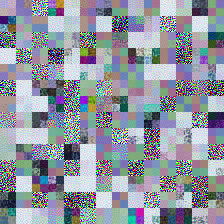} \\
\hline
$0.2$ &
\includegraphics[width=\unifiedsize]{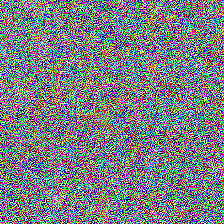} &
\includegraphics[width=\unifiedsize]{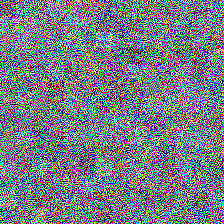} &
\includegraphics[width=\unifiedsize]{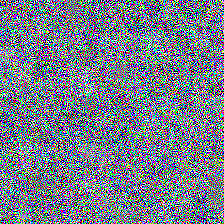} \\
\hline
$0.5$ &
\includegraphics[width=\unifiedsize]{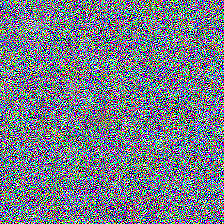} &
\includegraphics[width=\unifiedsize]{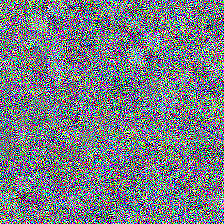} &
\includegraphics[width=\unifiedsize]{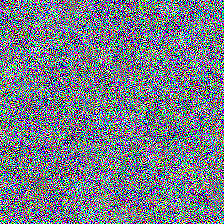} \\
\hline
$0.8$ &
\includegraphics[width=\unifiedsize]{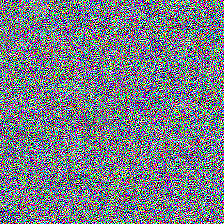} &
\includegraphics[width=\unifiedsize]{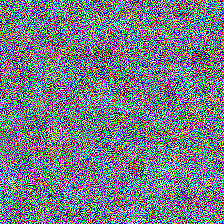} &
\includegraphics[width=\unifiedsize]{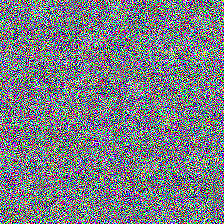} \\
\hline
\end{tabular}

\caption{Images restored by APRIL}
\label{fig:RW_RE_images}
\end{subfigure}

\caption{Effectiveness of FLRSP and image encryption against APRIL.}
\label{fig:APRIL_result}
\end{figure}

\begin{figure}[htb]
\centering
\newcommand{\unifiedsize}{2.59cm}
\newcommand{\imgmargin}{\hspace{0.5cm}}
\begin{subfigure}{\linewidth}
\centering
\begin{tabular}{cc}
\includegraphics[width=\unifiedsize]{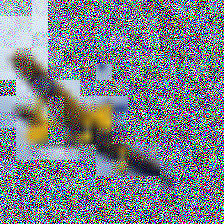} &
\imgmargin
\includegraphics[width=\unifiedsize]{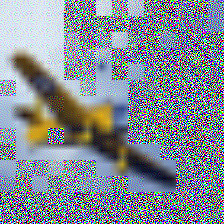} \\
{\small $R_W=10^{-4}$} &
\imgmargin
{\small $R_W=10^{-5}$}
\end{tabular}
\subcaption{$R_E=0.0$ (without image encryption)}
\label{fig:images_nonEnc_weak}
\end{subfigure}
\vspace{0.45em}

\begin{subfigure}{\linewidth}
\centering
\begin{tabular}{cc}
\includegraphics[width=\unifiedsize]{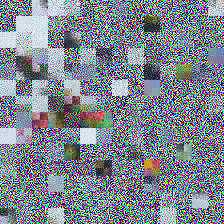} &
\imgmargin
\includegraphics[width=\unifiedsize]{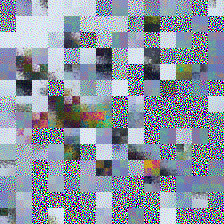} \\
{\small $R_W=10^{-4}$} &
\imgmargin
{\small $R_W=10^{-5}$}
\end{tabular}
\subcaption{$R_E=0.5$ (with partial image encryption)}
\label{fig:images_Enc_weak_05}
\end{subfigure}
\vspace{0.45em}

\begin{subfigure}{\linewidth}
\centering
\begin{tabular}{cc}
\includegraphics[width=\unifiedsize]{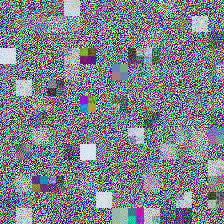} &
\imgmargin
\includegraphics[width=\unifiedsize]{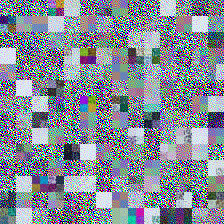} \\
{\small $R_W=10^{-4}$} &
\imgmargin
{\small $R_W=10^{-5}$}
\end{tabular}
\subcaption{$R_E=1.0$ (with full image encryption)}
\label{fig:images_Enc_weak}
\end{subfigure}
\caption{Images restored by APRIL with extremely small $R_W$.}
\label{fig:APRIL_RBW_weak_result}
\end{figure}
\par
Fig. \ref{fig:APRIL_result} illustrates an example of the attack results. Fig. \ref{fig:APRIL_result}\subref{fig:RW_RE_images} shows the restored results obtained by applying APRIL. As can be observed from the result with $R_W=0.0$ and $R_E=0.0$, the visual information of the training images was restored with high accuracy when neither FLRSP nor image encryption was applied.
\par
Next, the probability of generating zeros, $R_W$, was set to small values of $10^{-4}$ and $10^{-5}$. Under this condition, APRIL was conducted on the images shown in Figs. \ref{fig:APRIL_result}\subref{fig:img_original_line}, \subref{fig:img_05_original_line}, and \subref{fig:img_10_original_line}. Fig. \ref{fig:APRIL_RBW_weak_result} illustrates the restoration results. In the restoration of the image without image encryption shown in Fig. \ref{fig:APRIL_result}\subref{fig:img_original_line}, parts of the training image were successfully restored, as shown in Fig. \ref{fig:APRIL_RBW_weak_result}\subref{fig:images_nonEnc_weak}. For the image encrypted with $R_E=0.5$, the restored image contains regions visually similar to the original image, as shown in Fig.~\ref{fig:APRIL_RBW_weak_result}\subref{fig:images_Enc_weak_05}. In contrast, for the image in Fig. \ref{fig:APRIL_result}\subref{fig:img_10_original_line}, to which the proposed encryption was applied with $R_E=1.0$, the visual information of the training image remained sufficiently obfuscated even after the attack, as shown in Fig. \ref{fig:APRIL_RBW_weak_result}\subref{fig:images_Enc_weak}.

\par

These results show that the contribution of each technique depends on the parameter setting. When $R_W$ is relatively large, FLRSP strongly disturbs the reconstruction by APRIL. In contrast, the extremely small $R_W$ values in Fig. \ref{fig:APRIL_RBW_weak_result} were used to examine the effect of image encryption when gradient modification has only a limited effect. Under this condition, comparison between Figs. \ref{fig:APRIL_RBW_weak_result}\subref{fig:images_nonEnc_weak} and \ref{fig:APRIL_RBW_weak_result}\subref{fig:images_Enc_weak} shows that image encryption with $R_E=1.0$ reduces the visual interpretability of the reconstructed images. These observations suggest that the two techniques can play complementary roles, although their combination does not guarantee protection against all reconstruction attacks.

\subsection{Impact of Encryption on Image Classification}\label{Result_Acc}

\begin{table}[tb]
\centering
\caption{Impact of robustness enhancement using random binary weights and image encryption on classification accuracy (Bold indicates proposed method).}
\label{tab:result_model}
\begin{tabular}{c|ccc}
\diagbox{$R_W$}{$R_E$} 
 & $0.0$ & $0.5$ & $1.0$ \\
\hline
$0.0$& 98.44\% & 88.84\% & 51.36\% \\
$0.2$  & 98.38\% & \textbf{89.47\%} & \textbf{52.45\%} \\
$0.5$ & 98.37\% & \textbf{89.49\%} & \textbf{52.37\%} \\
$0.8$  & 98.32\% & \textbf{88.63\%} & \textbf{50.50\%} \\
\hline
\end{tabular}
\end{table}
To evaluate the impact of the proposed method on the classification performance of the model, image classification was conducted using the CIFAR-10 dataset. The 50,000 training images of CIFAR-10 were divided into five non-overlapping subsets of 10,000 images and assigned to each client. Training was conducted for 10 epochs using FedSGD with a batch size of 32 and a learning rate of 0.001. The present experiment focuses on the relative effects of $R_W$ and $R_E$ under a common training setting and does not fully evaluate long-term convergence or communication efficiency. In FedSGD, clients send gradients to the server for each batch, and the global model is updated. After training, the classification accuracy was evaluated on the 10,000 test images using the final global model.
\par
Table \ref{tab:result_model} shows the classification accuracy results. It can be observed that when only image encryption was applied, the classification accuracy decreased as the encryption strength $R_E$ increased. Specifically, $R_E = 0.5$ achieved a higher classification accuracy than $R_E = 1.0$; however, some visual information from the training image could be recovered under attack, as shown in Section~\ref{Attack}. These results indicate a trade-off between visual privacy and classification accuracy under the evaluated settings. In contrast, when only FLRSP was applied, the classification accuracy was almost maintained regardless of the probability of generating zeros, $R_W$. Furthermore, as indicated by the bold values in the table, the classification accuracy of the proposed method, which applies both FLRSP and image encryption, is comparable to that obtained when only image encryption is applied. These results demonstrate that applying both FLRSP and image encryption does not cause any additional degradation in classification performance. The classification accuracy for the extremely small values $R_W=10^{-4}$ and $10^{-5}$, which were used only for the reconstruction experiment in Fig. \ref{fig:APRIL_RBW_weak_result}, was not evaluated in this study and remains to be examined in future work.

\section{Conclusion}

In this paper, we proposed a security enhancement method based on dual obfuscation against image restoration attacks in federated learning. The proposed method combines FLRSP, which modifies shared gradients using random binary weights, with an image encryption technique that obfuscates the visual information of training images and allows independent keys for each client and each image.
\par
Experimental results using ViT showed that the proposed method reduces the visual information recovered by APRIL under the evaluated settings without causing additional classification degradation beyond that caused by image encryption. The results also showed a trade-off between visual privacy and classification accuracy depending on the encryption strength. Although the two techniques provide complementary protection, their combination does not guarantee protection against all reconstruction attacks.
\par
Future work will evaluate the proposed method with other federated learning methods and models, as well as against other and adaptive reconstruction attacks. We will also investigate its combination with differential privacy.

\end{document}